\RequirePackage{fix-cm}
\documentclass{svjour3}                     
\smartqed  
\usepackage{graphicx}
\usepackage[authoryear]{natbib}
\usepackage[utf8]{inputenc}
\usepackage{hyphenat}
\usepackage{threeparttable, tablefootnote}
\usepackage{ragged2e}
\usepackage{booktabs, tabularx}
\usepackage{todonotes}
\usepackage{subcaption}
\renewcommand{\cite}[2][]{\citep[#1]{#2}}
\journalname{ArXiv}
\begin{document}

\newcolumntype{L}{>{\raggedright\arraybackslash\hsize=.7\hsize}X}
\newcolumntype{m}{>{\raggedright\arraybackslash\hsize=.4\hsize}X}
\newcolumntype{n}{>{\raggedright\arraybackslash\hsize=.5\hsize}X}
\newcolumntype{o}{>{\raggedright\arraybackslash\hsize=.9\hsize}X}

\title{Visual Explanation for Identification of the Brain Bases for Dyslexia on fMRI Data}

\titlerunning{Visual Explanations for fMRI Data}        

\author{Laura Tomaz Da Silva        \and
        Nathalia Bianchini Esper \and 
        Duncan D. Ruiz \and
        Felipe Meneguzzi \and
        Augusto Buchweitz
}


\institute{L. Tomaz, D. D. Ruiz, F. Meneguzzi, N. B. Esper $*$, A. Buchweitz $**$\at
              PUCRS, School of Technology, Porto Alegre 90619-900, Rio Grande do Sul, Brazil \\
              $*,**$PUCRS, Graduate School of Medicine, Neurosciences, Porto Alegre 90619-900, Rio Grande do Sul, Brazil \\
              $*,**$BraIns, Brain Institute of Rio Grande do Sul, Porto Alegre 90619-900, Rio Grande do Sul, Brazil \\
              $*$ PUCRS, School of Health and Life Sciences, Psychology, Porto Alegre 90619-900, Rio Grande do Sul, Brazil \\
              \email{laura.tomaz@edu.pucrs.br, duncan.ruiz@pucrs.br, felipe.meneguzzi@pucrs.br, nathalia.esper@acad.pucrs.br, augusto.buchweitz@pucrs.br }  
}

\date{Received: date / Accepted: date}

\maketitle

\begin{abstract}
Brain imaging of mental health, neurodevelopmental and learning disorders has coupled with machine learning to identify patients based only on their brain activation, and ultimately identify features that generalize from smaller samples of data to larger ones. 
However, the success of machine learning classification algorithms on neurofunctional data has been limited to more homogeneous data sets of dozens of participants. 
More recently, larger brain imaging data sets have allowed for the application of deep learning techniques to classify brain states and clinical groups solely from neurofunctional features. 
Deep learning techniques provide helpful tools for classification in healthcare applications, including classification of structural 3D brain images. 
Recent approaches improved classification performance of larger functional brain imaging data sets, but they fail to provide diagnostic insights about the underlying conditions or provide an explanation from the neural features that informed the classification. 
We address this challenge by leveraging a number of network visualization techniques to show that, using such techniques in convolutional neural network layers responsible for learning high-level features, we are able to provide meaningful images for expert-backed insights into the condition being classified. 
Our results show not only accurate classification of developmental dyslexia from the brain imaging alone, but also provide automatic visualizations of the features involved that match contemporary neuroscientific knowledge, indicating that the visual explanations do help in unveiling the neurological bases of the disorder being classified.
\keywords{Visual Explanation \and Deep Learning \and Dyslexia \and Neuroimaging \and fMRI}
\end{abstract}

\section{Introduction} \label{intro}


Brain imaging techniques such as structural MRI, functional MRI (fMRI) and diffusion-weighted imaging (DWI),
can be  used to find  altered cortical tissue, structure and function associated with mental health disorders~\cite{atluri2013complex}. 
These techniques allow for the identification of neural markers, which in turn may provide or inform a diagnosis based on image features~\cite{american2013diagnostic}.

Recent advances in deep learning have led researchers to employ machine learning to automate the analysis of medical imaging, including neurological images~\cite{craddock2009disease, tamboer2016machine, froehlich2014classifying}. 
The most successful technique derived from deep learning for image classification consists of building neural network with convolutional layers, i.e. Convolutional Neural Networks (CNNs). The CNN specializes in processing multiple arrays, such as images (2D), audio and video or volumetric data (3D)~\cite{bengio2015deep}.

Brain imaging volumes have tens of thousands of voxels (3D\hyp{pixel}) per image. Neurofunctional indices are mapped to these voxels, which makes feature selection a challenge for most machine learning approaches. Supervised approaches to machine learning relied on experts for feature selection~\cite{bengio2015deep}. Deep learning approaches obviate the dependence on supervision by automatically learning the features that better represent the problem domain~\cite{bengio2015deep}.
Before deep learning methods were effectively applied to classification of brain imaging data, support vector machine (SVM) learning algorithms was the frequent choice for machine learning analyses of brain imaging~\cite{cortes1995support}. SVM algorithms have the ability to generalize well in smaller fMRI datasets~\cite{just2017machine, li2014,Buchweitz2012, murphy2012machine, craddock2009disease, tamboer2016machine, froehlich2014classifying}, which are typically in the dozens of participants due to the high costs of fMRI scans~\cite{craddock2009disease, froehlich2014classifying}. 
Moreover, SVM models trained with linear kernels offer relatively straightforward explanations. 
This SVM property may be useful to break the ``curse of dimensionality'' by reducing the risk of overfitting the training data. The number of voxels used in feature selection should be reduced as much as possible. 

Feature selection for brain imaging data is often performed on voxels in anatomically or functionally defined regions\hyp{of}\hyp{interest} (ROIs) based on the literature ~\cite{wolfers2015estimating} or by data\hyp{driven} methods that establish clusters of stable voxels~\cite{just2014identifying, shinkareva2008using}. 
By contrast, deep learning models learn feature hierarchies at several levels of abstraction, which allows the system to learn complex functions independent of human-crafted features~\cite{bengio2015deep}. 
CNNs are applicable to a variety of medical image analysis problems, such as disorder classification~\cite{heinsfeld2018identification}, anatomy or tumor segmentation~\cite{kamnitsas2017efficient}, lesion detection and classification~\cite{ghafoorian2017deep}, survival prediction~\cite{van2017deep}, and medical image construction~\cite{li2014}. 
Although these models can be accurate, their conclusions are opaque to human understanding and lack a straightforward explanation to help  diagnosis. 
It is thus difficult for healthcare practitioners to apply and trust the results of machine learning models of brain imaging to assist them in their clinical diagnoses. Providing accurate visual representation of neural networks involved in deep learning classification may be a step in the direction of improving diagnostic application of classification using neurofunctional indices.

The goal of the present study is to employ feature visualization techniques for CNNs. 
These techniques produce visual explanations of the key brain regions used to classify patients based solely on brain function. 
The key contribution is a visual representation of the regions involved in classifying whether children are dyslexic or not. 
The present technique provides a better understanding of CNN behavior and may provide practitioners with a tool to glean neural alterations associated with a disorder from functional brain imaging scans.

\section{Method}\label{sec:method}

\subsection{Data}\label{sec:data}
The brain imaging data was collected as part of a research initiative to investigate the neural underpinnings of dyslexic children in Brazil. 
The participants were diagnosed with dyslexia following a multidisciplinary evaluation that included medical history, reading and writing tests~\cite{Toazza2017, Costa2015}, and an IQ test (Wechsler Abbreviated Scale of Intelligence~\cite{wechsler2012wechsler}). The dyslexics readers were evaluated at as part of an umbrella project that aims to establish a brain database of dyslexic readers of Brazilian Portuguese~\cite{buchweitz2019decoupling, Costa2015}. The typical readers were part of a longitudinal investigation of children learning to read~\cite{buchweitz2019decoupling}. 

\subsubsection{Participants}
The present study included 32 children who were divided into two groups: typical readers (TYP; n = 16) and dyslexic readers (DYS; n = 16)~\cite{buchweitz2019decoupling}. The participants were all monolingual speakers of Portuguese and right\hyp{handed}. The two groups were matched for age, sex and IQ [age \(7\)$-$\(13\) (\(9\) $\pm$ \(1.39\))]. The typical readers children were evaluated at the end of the 2014 school year, and were scanned during the 2015 school year. The 16 dyslexic children were scanned between 2014 and 2015~\cite{buchweitz2019decoupling}. Table~\ref{table:demographics} summarizes the complete demographics on this dataset.

\begin{table}
\centering
\begin{tabular*}{0.8\textwidth}{l l l}
\toprule
Groups & Typical Readers & Dyslexics \\
\midrule
No. of Subjects & 16 & 16 \\
Age (mean $\pm$ STD) & 8.44 $\pm$ 0.51 & 9.63 $\pm$ 0.88 \\
Sex (Male / Female) & 09 / 07 & 11 / 05 \\
\bottomrule
\end{tabular*}
\caption{Demographic information of children included in the study.}\label{table:demographics}
\end{table}

\subsubsection{Word-reading task}

Task based fMRI examines brain regions whose activity changes from baseline in response to the performance of a task or stimulus~\cite{petersen2012mixed}. The study was designed as a mixed event-related experiment using a word and pseudoword reading test validated for Brazilian children~\cite{salles2013normas}. The task consisted of 20 regular words, 20 irregular words, and 20 pseudowords. The 60 stimuli were divided into two 30-item runs to give the participants a break halfway into the task. Words and pseudowords were presented on the screen one at a time for 7 seconds each. A question was presented to participants along with each word ("Does the word exist?"), to which participants had to select "Yes" or "No" by pressing response buttons. After 10 trials (10 words) either a baseline condition or rest period was inserted in the experimental paradigm. The baseline condition consisted of presentation of a plus sign "+" in the middle of the screen for 30 seconds, during which participants were instructed to relax and clear their minds.

\subsubsection{Data acquisition}

Data was collected on a GE HDxT 3.0 T MRI scanner with an 8-channel head coil~\cite{buchweitz2019decoupling}. The following MRI sequences were acquired: a T1 structural scan (TR/TE = 6.16/2.18 ms, isotropic 1mm\textsuperscript{3} voxels); two task-related 5-min 26-sec functional FMRI EPI sequences; and a 7\hyp{min} resting state sequence. The task and the resting\hyp{state} EPI sequences used the following parameters: TR = 2000 ms, TE = 30 ms, 29 interleaved slices, slice thickness = 3.5 mm; slice gap = 0.1 mm; matrix size = 64 $\times$ 64, FOV = 220 $\times$ 220 mm, voxel size = 3.44 $\times$ 3.44 $\times$ 3.60 mm~\cite{buchweitz2019decoupling}.

\subsubsection{Data preprocessing}

The preprocessing steps for the task\hyp{based} (word\hyp{reading} task) fMRI are described as follows. 
Word\hyp{reading} task: preprocessing included slice\hyp{time} and motion correction, smoothing with a 6mm FWHM Gaussian kernel, and a nonlinear spatial normalization to 3.0 $\times$ 3.0 $\times$ 3.0 mm voxel template (HaskinsPedsNL template). TRs with motion outliers (\textgreater0.9 mm) were censored from the data. The criteria for exclusion due to head motion were: excessive motion in 20\% of the TRs. The average head motion for each group for the participants included in the study, in the word\hyp{reading} paradigm, was: DYS M = 0.16 $\pm$ 0.08, TYP M = 0.18 $\pm$ 0.15. One participant from each group was excluded due to excessive head motion~\cite{buchweitz2019decoupling}.

\subsection{Classification Task}

We trained a number of deep learning models for the classification task using two key recent techniques in learning for image classification: CNNs~\cite{lecun1998gradient} and data augmentation~\cite{perez2017effectiveness}.
For the CNNs, we evaluated both two-dimensional (2D) and three-dimensional (3D) CNNs. 

\begin{figure}
\centering
\includegraphics[height=3cm]{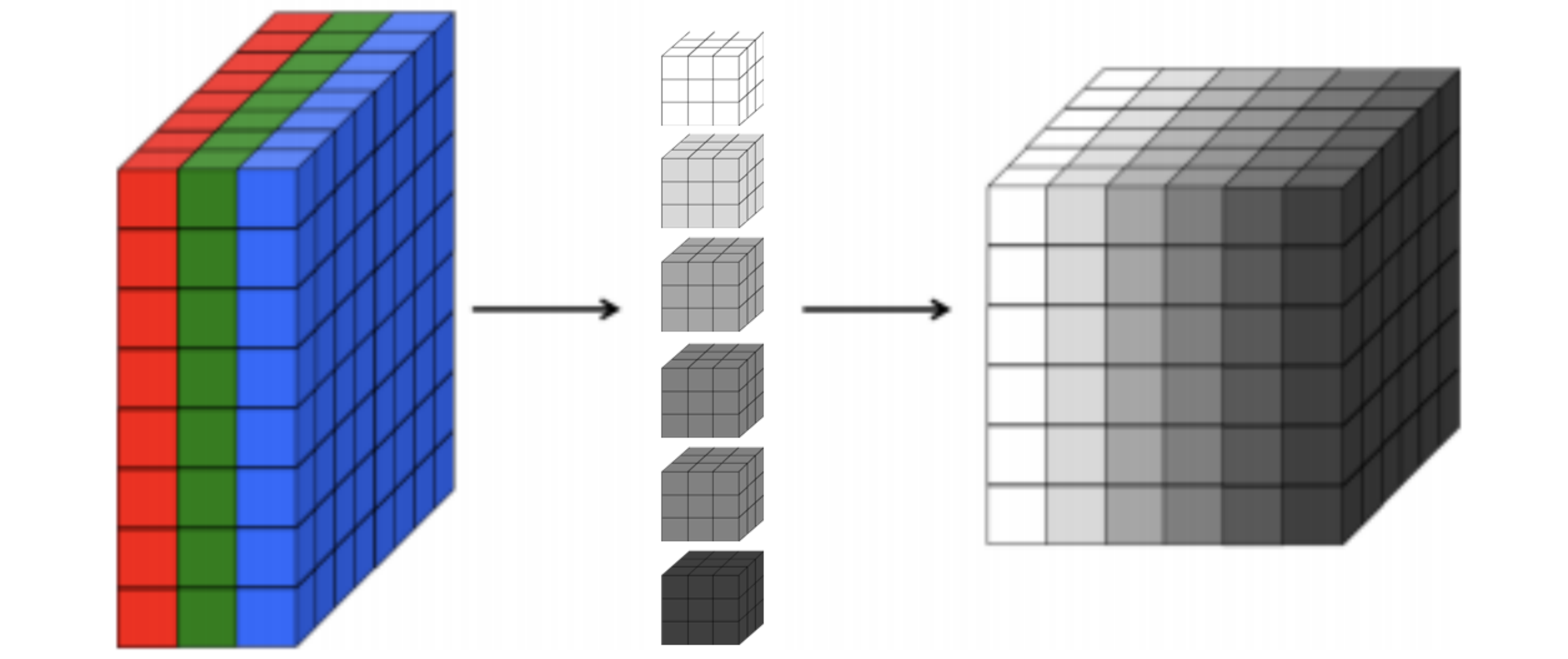}
\caption{\label{fig:convolution} A three-dimensional representation of a convolutional layer, each filter represents a slice in the output.}
\end{figure}

First, regarding the CNNs, our choice of model focuses on 2D CNNs due to the size of our dataset.
Specifically, 2D CNNs have a smaller number of parameters in comparison to 3D CNNs~\cite{szegedy2015going}. 
Thus, training a 3D CNN necessitates substantially larger datasets in order to generalize well. 
Indeed, our experiments show that 2D CNNs achieve superior accuracy to 3D CNNs given the limitations of our dataset. 
A 2D CNN takes an input having three dimensions (a height \(h\), a width \(w\), and a number of color channels or a depth \(d\)). 
This input volume is then processed by \(k\) filters, which operate on the entire volume of feature maps that have been generated at a particular layer. 
2D convolutions have a pseudo third dimension comprising the color channels in each image, such that a 2D CNN applies convolutions to each channel separately, combining the resulting activations. 
Figure~\ref{fig:convolution} illustrates each RGB channel in the input as a slice. 
A filter, which corresponds to weights in the convolutional layer, is then multiplied with a local portion of the input to produce a neuron in the next volumetric layer of neurons. 
In the Figure~\ref{fig:convolution}, the middle part represents filters, the depth of the filter corresponds to the depth of the input. 
The last cube in the figure represents the output activations of the combined convolution operations for each channel.
The depth of the output volume of a convolutional layer is equivalent to the number of filters in that layer, that is, each filter produces its own slice. 
This can be viewed as using a 3D convolution for each output channel, which happens to have the same depth as the input~\cite{buduma2017fundamentals}. 
For this reason, it is possible to use volumetric images as inputs to a 2D CNN. 
In effect, this means that a 2D CNN processes the 3D volume of brain scan activations slice-by-slice. 

Second, we avoid overfitting in our small dataset by employing data augmentation. 
Data augmentation is a technique~\cite{perez2017effectiveness} that provides the model with more data to increase the model's ability to generalize from it. 
Such techniques are already employed in several image problems in deep learning models, but are still incipient in fMRI data~\cite{Grochowski2018dataaug}. 

We adopted two approaches to build the 2D CNN architectures: i) use genetic programming, more specifically grammar-based genetic programming (GGP) fitted to our problem; and ii) employ a modified version of the LeNet-5~\cite{lecun1998gradient} classification model. 
We then trained the resulting architecture using our dataset, and compared the effectiveness of 3D convolutions by converting the generated 2D CNNs into 3D ones by swapping the 2D convolutional layers to appropriately-sized 3D convolutions. 

\subsection{Visual Explanations Task}

While many application areas for machine learning focus simply on model performance, recent work has highlighted the need for explanations for the decisions of trained models. 
Most users of machine learning often want to understand the trained models in order to gain confidence in the predictions. 
This is especially true for machine learning models used in medical applications, where the consequences of each decision must be carefully explained to patients and other stakeholders~\cite{yang2018visual, jin2019attention}.
Besides the explainability aspect required of direct medical applications, our key motivation is to allow neuroimaging specialists to derive new insights on underpinnings of specific learning disorders such as dyslexia. 
Indeed, clinical diagnosis of dyslexia is reliable and costs less than using fMRI scans to validate such diagnostics~\cite{torgesen1998catch, ramus2003relationship}. 
However, researchers of dyslexia are interested in further understanding the disorder and its neural underpinnings \emph{in\hyp{vivo}}~\cite{shaywitz2001neurobiology, hoeft2011neural}. 
For this reason, building data-driven diagnostics models via machine learning and generating explanations for such models can be an invaluable tool for dyslexia research. 

Recent research developed several methods for understanding and visualizing CNNs, in part as a response to criticism that the learned features in a neural network are not interpretable to humans~\cite{ZeilerF13, szegedy2013intriguing, zhou2016learning}. 
A category of techniques that aim to help understand which parts of an image a CNN model uses to infer class labels is called Class Activation Mapping \textsc{(CAM)}~\cite{zhou2016learning}. 
\textsc{CAM} produces heatmaps of class activations over input images. 
A class activation heatmap is a 2D grid of scores associated with a particular output class, computed for every location for an input image, indicating how important each location is with respect to that output class~\cite{zhou2016learning}. 
CAM can be used by a restricted class of image classification CNNs, precluding the model from containing any fully-connected layers and employing global average pooling (GAP).

A recent approach to visualize features learned by a CNN is \textsc{Grad-CAM}~\cite{selvaraju2017grad}. \textsc{Grad-CAM} is a generalization of \textsc{CAM} and can be applied to a broader range of CNN models without the need to change their architecture.
Instead of trying to propagate back the gradients, \textsc{Grad-CAM} infers a downsampled relevance heatmap of the input pixels from the activation heatmaps of the final convolutional layer. 
The downsampled heatmap is upsampled to obtain a coarse relevance heatmap. 
This approach has two key advantages: first, it can be applied to any CNN architecture; and second, it requires no re-training or change in the existing neural network architecture.

\begin{figure*}[t]
    \centering
    \includegraphics[width=.8\textwidth,height=5cm]{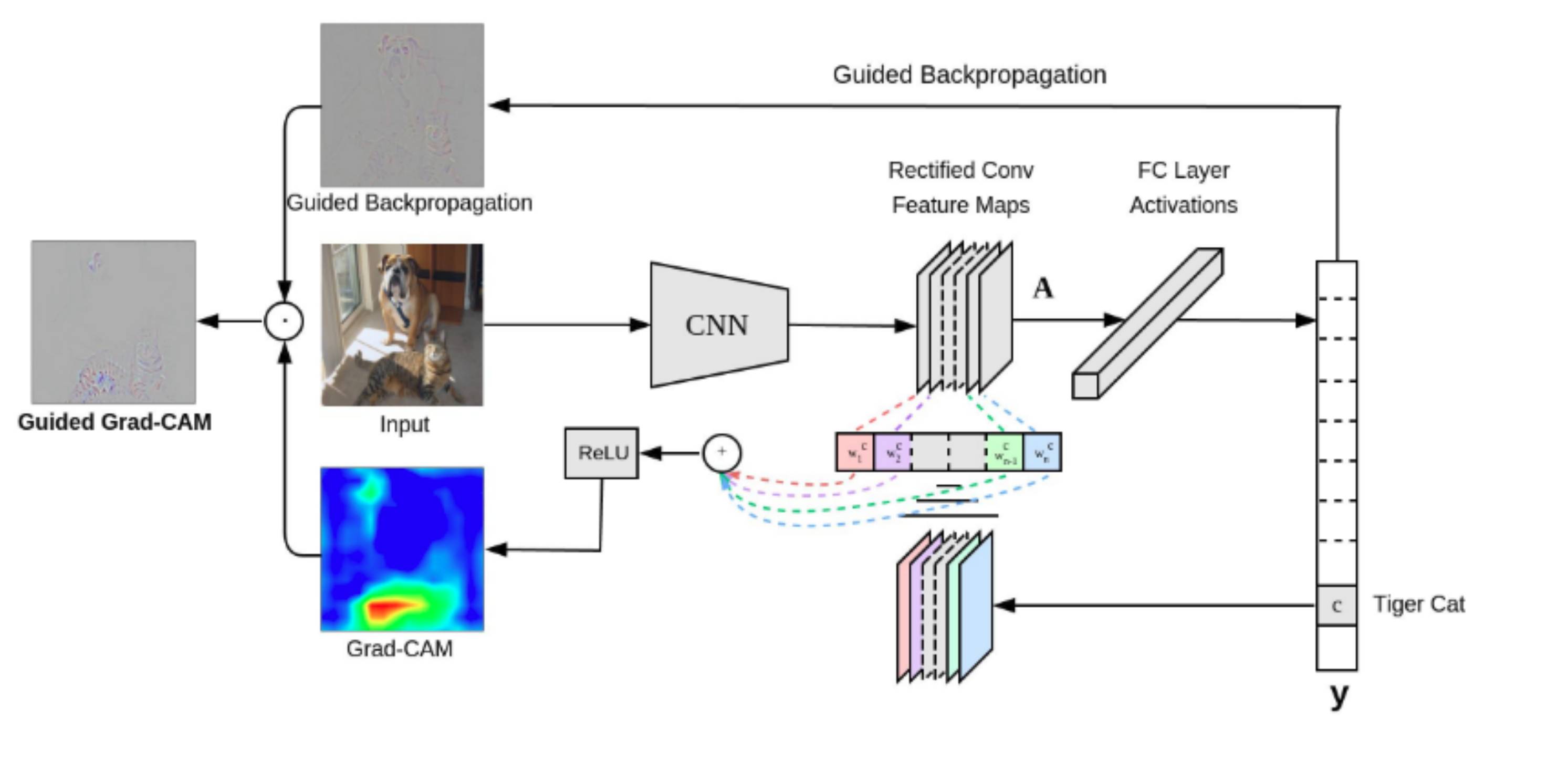}
    \caption{\textsc{Grad-CAM} overview~\cite{selvaraju2017grad}.}
    \label{fig:gradcam}
\end{figure*}

Figure~\ref{fig:gradcam} illustrates the \textsc{Grad-CAM} approach. 
Given an image and a class of interest (in the example, 'tiger cat') as input, \textsc{Grad-CAM} first forward propagates the image through the CNN part of the model and then through task-specific computations to obtain a raw score for the category. 
Next, \textsc{Grad-CAM} sets all the gradients that do not belong to the desired class (tiger cat), which are originally set to one, are set to zero. 
\textsc{Grad-CAM} then backpropagates this signal to the rectified convolutional feature maps of interest, which it combines to compute the coarse \textsc{Grad-CAM} localization (the blue heatmap in the figure) which represents where the model has to look to make the particular decision. 
Finally, \textsc{Grad-CAM} pointwise multiplies the heatmap with guided backpropagation to get Guided \textsc{Grad-CAM} visualizations which are both high-resolution and concept-specific~\cite{selvaraju2017grad}.

\section{Experiments and Results}

\subsection{Classification}

The deep learning classification model was implemented using the Keras open source library~\cite{chollet2015keras} and trained with an Nvidia Geforce GTX 1080 Ti graphical processing unit (GPU) with 12 GB of memory.
In our genetic programming (GP) approach, we generated a population of CNN architectures, such that each CNN architecture was an individual in a population, and which was evaluated to produce a fitness value. 
Network topology for all CNNs generated was based on a specific grammar for our problem and a set of different hyperparameters.

\begin{figure*}[t]
    \centering
    \includegraphics[width=\textwidth,height=5cm]{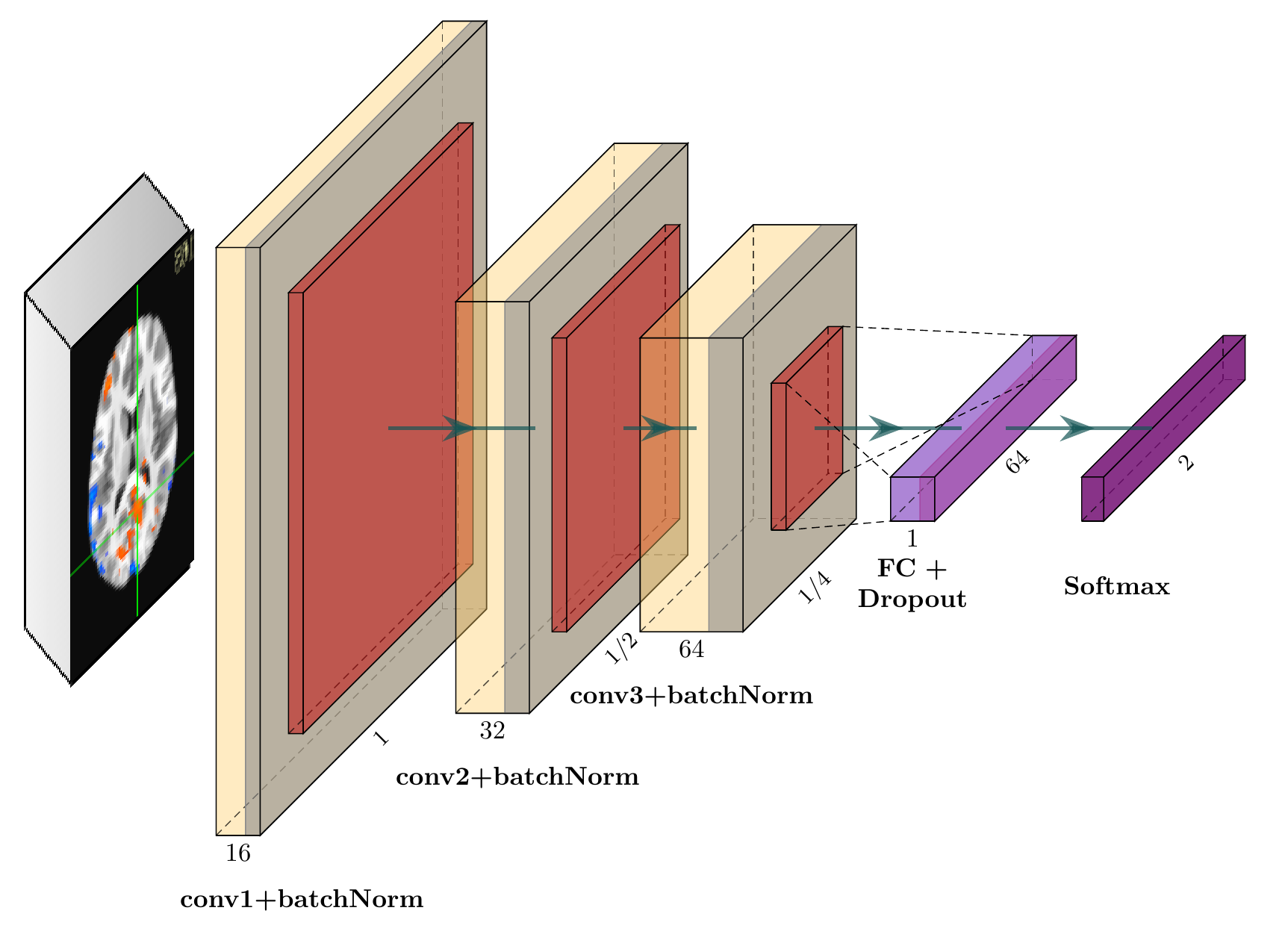}
    \caption{Modified LeNet-5 overview}
    \label{fig:lenet5}
\end{figure*}

We introduced four key modifications in our version of the LeNet\hyp{5} architecture. 
First, we added batch normalization layers in the convolutional layers to improve convergence and generalization~\cite{ioffe2015batch}. 
Second, we used \textsc{ReLU} activations in the convolutional layers instead of $\tanh$. 
Third, we changed the average pooling to max pooling in the subsampling layers. 
Finally, we used a dropout rate of \(0.5\) in the fully connected layer. 
Figure~\ref{fig:lenet5} illustrates our modified version of LeNet\hyp{5}. 
Our model architecture contains approximately \(175 k\) parameters, a small amount in comparison to deeper architectures, such as VGG\hyp{16}~\cite{simonyan2014very}, which contains over \(138\) million parameters.

Our 3D CNN was developed based on our 2D CNN model. 
We made the changes necessary to adapt 2D convolutions, 2D pooling layers to a 3D model. 
In order to fit our data to a 3D CNN model, we expanded our data adding one channel for gray images resulting in a 4-dimensional array as input to the network.
The resulting architecture has over 3 million parameters.

We compared our induced deep learning models with the (SVMs)~\cite{cortes1995support} technique, which has been used in a substantial number of previous neuroimaging studies~\cite{tamboer2016machine, froehlich2014classifying}. 
Specifically, this technique is popular for fMRI applications because datasets typically have many features (voxels), but only a relatively small set of subjects. 

We trained all models to classify the participants between dyslexics and typical readers using the Adam optimizer. 
We improved the performance of our classifier by employing two data augmentations to our dataset: i) we added Gaussian noise to fMRI images to generalize to noisy images; and ii) we added a random Gaussian offset, or contrast, to increase differences between images.
The input of our machine and deep learning models was the whole brain volume (\(60\) $\times$ \(73\) $\times$ \(60\) voxels) and a binary mask filling the brain volume to retrieve data from all brain regions. 
We split the dataset into train \(80\)\%, validation \(10\)\%, and test \(10\)\% sets.
The parameter values including learning rate, dropout rate, batch size, and epoch size were optimized using the ranges summarized in Table~\ref{tab:defaultConfigCNN}. 
Note that we optimized the batch size to use the maximum available GPU memory. 

\begin{table}
\centering
\begin{tabular}{l l}
\toprule [2pt]
Hyperparameters & Values\\
\hline
Kernel size & Ranging from \(1\) to \(5\) \\
\# of filters & Starts with \(16\); duplicates after every convolution\\
Stride & Ranging from \(1\) to \(3\) \\
Learning rate & Logarithmic range of [\(1\), \(0.1\), \(0.01\), \(0.001\), \(0.0001\), \(0.00001\)]  \\
Dropout Rate & Tuned in the range of [\(0.1\), \(0.5\), \(1\)]\\
Batch size & 16 \\
\# of epochs & Tuned in the range of [\(10\), \(50\), \(100\)] \\
\# of Neurons FC layer & Tuned in the range of [\(32\), \(64\), \(128\), \(256\), \(512\)] \\
\bottomrule [2pt]
\end{tabular}
\caption{CNN hyperparameters}
\label{tab:defaultConfigCNN}
\end{table}

All hyperparameters were optimized for both the 2D and 3D CNN models. For our SVM models, first, we applied an exhaustive search over specified parameters values for our SVM estimator. 
Second, we evaluated different methods of cross-validation. We report the results from splitting the data into train, validation, and test for Linear SVM implemented using scikit\hyp{learn}~\cite{scikit-learn} library in Python.

Our modified version of LeNet\hyp{5} 2D CNN network achieved \(85.71\)\% accuracy on subject classification. 
Our best GP 2D CNN model achieved an accuracy of \(94.83\%\) on subject classification.
In comparison to the 2D CNN architecture, the 3D CNN, from both the modified LeNet\hyp{5} and GP approach, had an inferior accuracy on subject classification. 
The 3D CNN was also more prone to overfitting in the first few epochs of training. 
By contrast, the SVM approach achieved much lower classification accuracy, regardless of the training dataset composition.
Table~\ref{table:results} summarizes the results from all our classification approaches. 

\begin{table}
\centering
\begin{tabularx}{0.8\textwidth}{l l}
\toprule
Technique & Accuracy  \\
\midrule
\textbf{Best GP 2D CNN} &  \textbf{\(94.83\%\)}\\
Modified LeNet-5 & \(85.71\%\) \\
Best GP 3D CNN & \(78.57\%\)\\
Modified LeNet-5 3D & \(71.43\%\) \\
SVM (\(80\)\% train, \(10\)\% validation, \(10\)\% test) & \(70\%\)\\
\bottomrule
\end{tabularx}
\caption{Summary of Dyslexia Classification Results.}\label{table:results}
\end{table}

\subsubsection{Visual Explanations}

After training the 2D CNN model, we loaded the model with the best accuracy to visualize the learned gradients using \textsc{Grad-CAM} technique~\cite{selvaraju2017grad}. 
The class activation generated by \textsc{Grad-CAM} shows which  regions were more instrumental to the classification. 

\begin{figure*}[t]
\centering
\begin{subfigure}[b]{.45\linewidth}
    \centering
    \includegraphics[width=\textwidth]{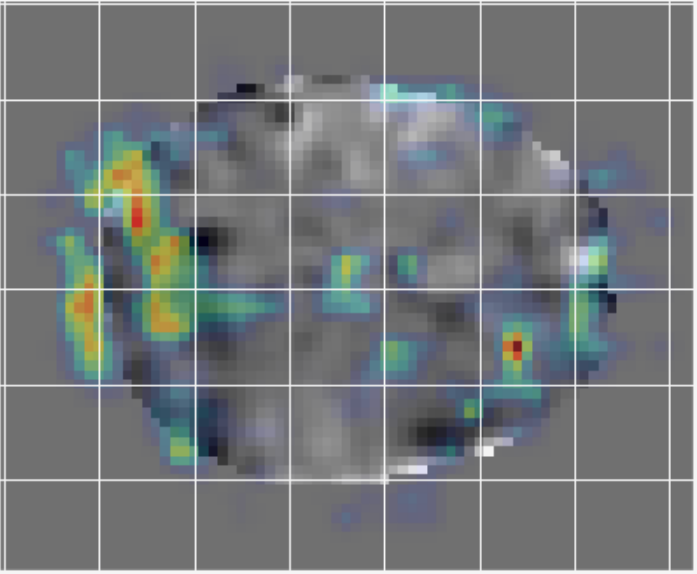}
    \caption{\label{fig:grad-ambac} Class activation mapping for dyslexic participants classification. 
    }
\end{subfigure}
\begin{subfigure}[b]{.45\linewidth}
    \centering
    \includegraphics[width=\textwidth]{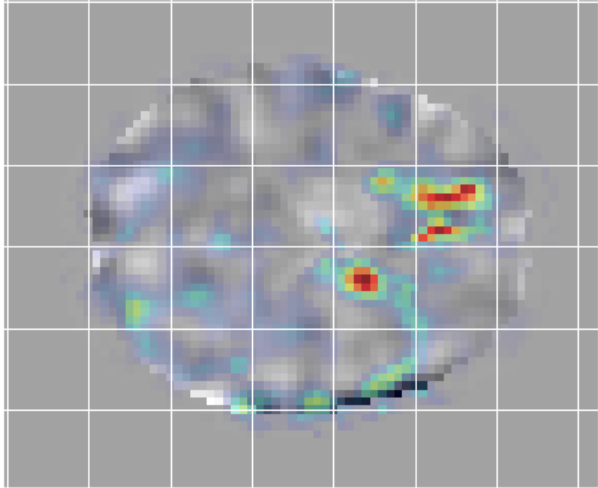}
    \caption{\label{fig:grad-schools} Class activation mapping for non-dyslexic participants classification. 
    }
\end{subfigure}
\caption{\textsc{Grad-CAM} technique.}
\end{figure*}

To employ \textsc{Grad-CAM} visualization to identify key differences between subjects and controls, we chose a pair of subjects as input, i.e. a control (non\hyp{dyslexic}) subject and a dyslexic subject to generate the class activation mappings. 
Figures~\ref{fig:grad-ambac} and~\ref{fig:grad-schools} show \textsc{Grad-CAM} generated images of control and dyslexic subjects, with respect to the gradients learned by the network model. 
Both images depict the central slice from the axial view of the brain volume. 
Areas with lower class activation mappings are colored in gray, whereas areas with higher class activation mappings are color-coded from yellow (instrumental) to red (more instrumental). 
The color coding thus represents the brain regions impact on the model classification of subjects. 

\begin{figure*}[t]
\centering
\begin{subfigure}[b]{.35\linewidth}
    \centering
    \includegraphics[width=\textwidth]{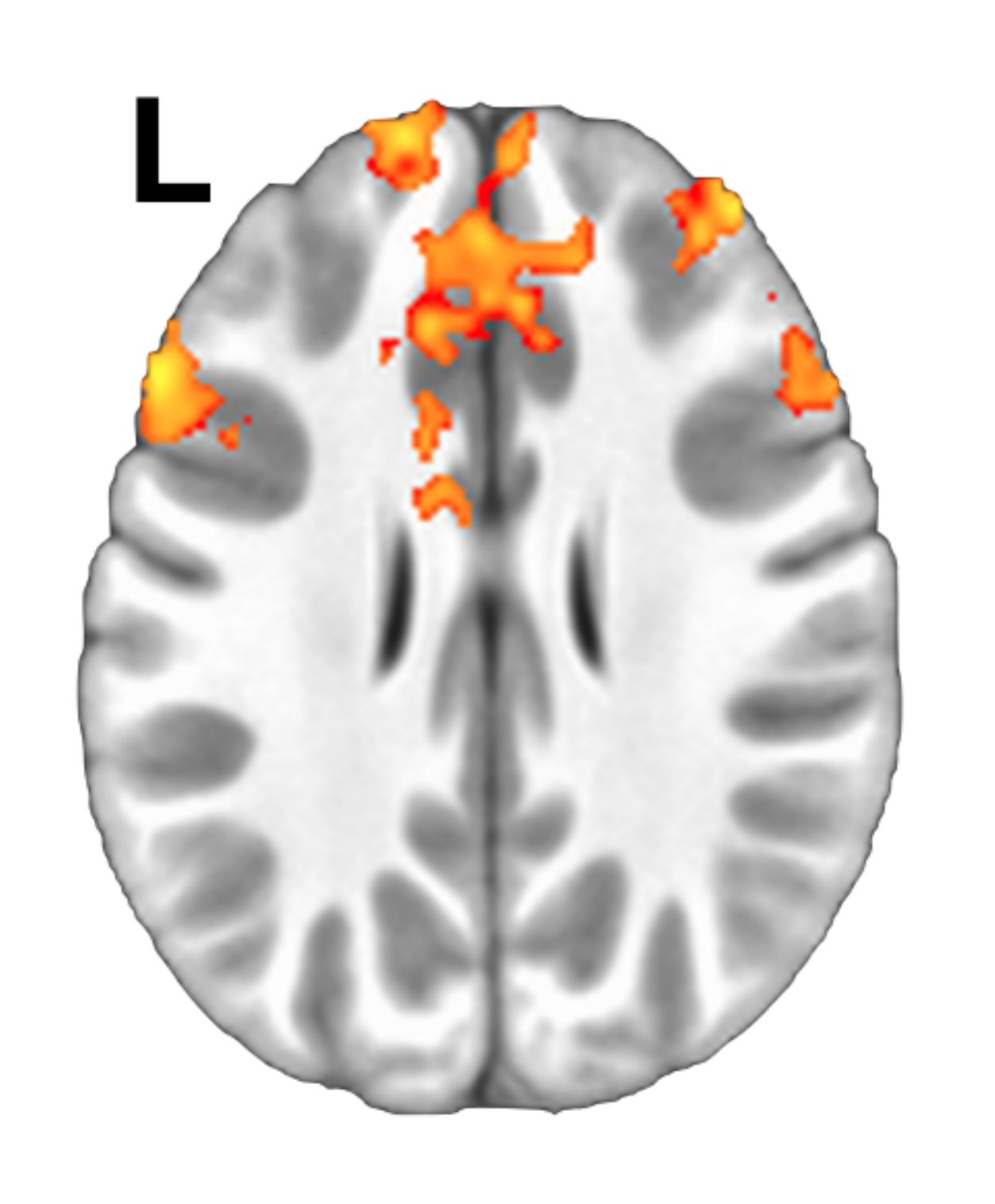}
    \caption{\label{fig:schools-visual} Example visual explanation for Control subjects. Activation highlights ACC. Image depicts a slice at $z$ = \(28\).}
\end{subfigure}
\begin{subfigure}[b]{.38\linewidth}
    \centering
    \includegraphics[width=\textwidth]{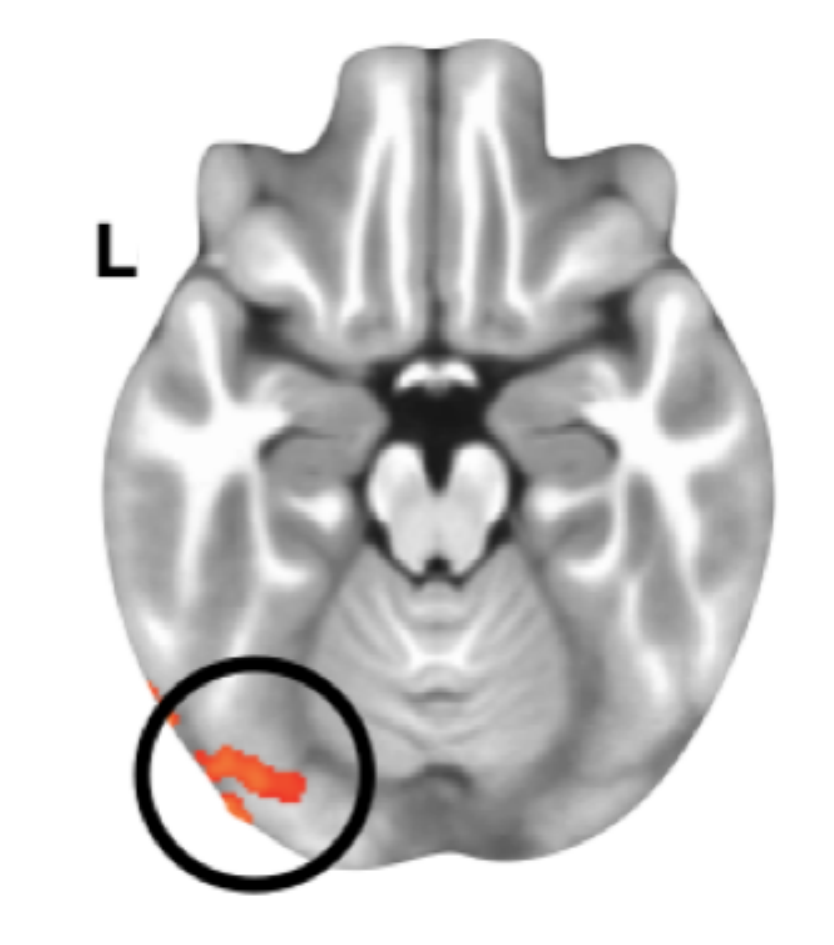}
    \caption{\label{fig:ambac-visual} Example visual explanation for Dyslexic subjects. Circle highlights left occipitotemporal region. Image depicts a slice at $z$ = \(-12\).}
\end{subfigure}
\caption{AFNI~\cite{Cox1996} images showing brain activation from \textsc{Grad-CAM}.}
\end{figure*}

The visualization showed regions that were instrumental to the classification. The regions included: (i) the left occipital lobe (including left fusiform gyrus) with a high classification mapping for dyslexic participants; (ii) the anterior cingulate cortex (ACC) with a high classification mapping for typical readers (controls). Figure~\ref{fig:ambac-visual} illustrates the classification mapping for dyslexia in left occipitotemporal region corroborates brain imaging findings that show functional alterations in this region associated with dyslexia and poor reading~\cite{shaywitz2002disruption, Martin2015}. Figure~\ref{fig:schools-visual}  illustrates high classification mappings found in ACC for controls; activation of the ACC is usually associated with strategic control and attention processes~\cite{Chein2005, Bush1999}. More ACC activation has been found in association with additional attention processes engaged by early good readers, and in association with poor readers who benefited the most from reading remediation in~\cite{shaywitz2002disruption, buchweitz2019decoupling, Shaywitz2003}. Other regions with high classification mapping for dyslexia and controls are shown in Table~\ref{tab:brainregions}.

\begin{table*}
\begin{tabularx}{0.698\textwidth}{|l|l|l|l|}
\hline
Dyslexic  &  & \textbf{Regions} & \textbf{$\#$ voxels} \\ \hline
Figure~\ref{fig:dysvisual}$a$ & Left & Inferior Parietal & $55$ \\ 
& & Postcentral & $72$ \\ 
& & Precentral & $9$ \\
& & Precuneus & $4$ \\
& & Superior Parietal & $24$ \\
& & Supramarginal & $26$ \\ \cline{1-4}
Figure~\ref{fig:dysvisual}$b$ & Left & Paracentral & $2$ \\ 
& & Postcentral & $3$ \\ 
& & Precentral & $7$ \\ 
& & Superior Frontal & $12$ \\\cline{2-4} 
& Right & Caudal Anterior Cingulate & $18$ \\ 
& & Inferior Parietal & $73$ \\
& & Paracentral & $1$ \\
& & Postcentral & $63$ \\
& & Precentral & $49$ \\
& & Precuneus & $25$ \\ 
& & Superior Frontal & $10$ \\ 
& & Superior Parietal & $19$ \\
& & Supramarginal & $41$ \\ \cline{1-4}  
Figure~\ref{fig:dysvisual}$c$ & Left & Superior Frontal & $28$ \\ \cline{2-4} 
& Right & Superior Parietal & $13$ \\ \cline{1-4}
Figure~\ref{fig:dysvisual}$d$ & Left & Caudal Anterior Cingulate & $19$ \\ 
& & Caudal Middle Frontal & $40$ \\ 
& & Posterior Cingulate & $3$ \\
& & Precentral & $26$ \\
& & Rostral Middle Frontal & $21$ \\
& & Superior Frontal & $22$ \\ \cline{2-4} 
& Right & Caudal Anterior Cingulate & $13$ \\ 
& & Caudal Middle Frontal & $53$ \\
& & Posterior Cingulate & $2$ \\
& & Precentral & $10$ \\
& & Superior Frontal & $49$ \\ 
& & Superior Parietal & $13$ \\ \cline{1-4}
\end{tabularx}
\caption{Voxel count per brain region of Dyslexics for Figure~\ref{fig:dysvisual}. Brain Regions Instrumental for Dyslexic Identification with \textsc{Grad-CAM}~\cite{selvaraju2017grad}. Region labels follow Haskins pediatric atlas~\cite{molfese2015haskins}.}
\label{tab:voxelcountdys}
\end{table*}

\begin{table*}
\begin{tabularx}{0.767\textwidth}{|l|l|l|l|}
\hline
Typical Readers  &  & \textbf{Regions} & \textbf{$\#$ voxels} \\ \hline
Figure~\ref{fig:typvisual}$a$ & Left & Caudal Middle Frontal & $2$ \\ 
& & Pars Opercularis & $35$ \\ 
& & Precentral & $2$ \\
& & Rostral Middle Frontal & $12$ \\
& Right & Inferior Parietal & $3$ \\ \cline{1-4}
Figure~\ref{fig:typvisual}$b$ & Left & Caudal Anterior Cingulate & $22$ \\ 
& & Posterior Cingulate & $1$ \\ 
& & Precentral & $7$ \\ 
& & Rostral Middle Frontal & $8$ \\
& & Superior Frontal & $59$ \\\cline{2-4} 
& Right & Caudate & $3$ \\ 
& & Caudal Anterior Cingulate & $6$ \\
& & Superior Frontal & $33$ \\
& & Postcentral & $63$ \\
& & Superior Parietal & $20$ \\ \cline{1-4}
\end{tabularx}
\caption{Voxel count per brain region of Typical readers for Figure~\ref{fig:typvisual}. Brain Regions Instrumental for Typical Readers Identification with \textsc{Grad-CAM}~\cite{selvaraju2017grad}. Region labels follow Haskins pediatric atlas~\cite{molfese2015haskins}.}
\label{tab:voxelcounttyp}
\end{table*}

\begin{figure*}[t]
    \centering
    \includegraphics[width=\textwidth]{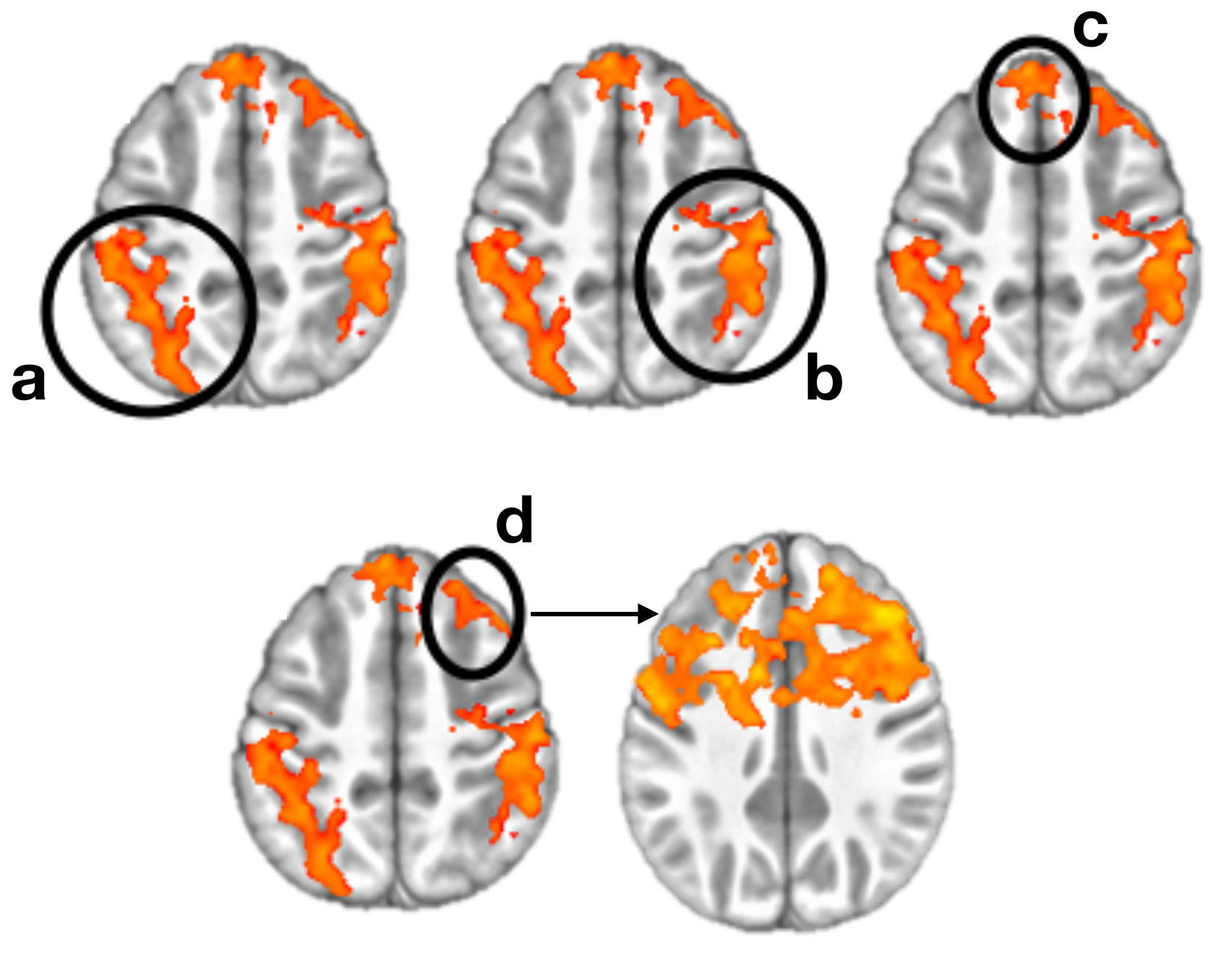}
    \caption{Example visual explanation for Dyslexic subjects. Circle highlights instrumental brain regions for Dyslexic identification summarized in Table~\ref{tab:voxelcountdys}. The left side of the images represent the left side of the brain. Images $a, b, c, d$ depict a slice at $z$ = \(48\) and the last image depicts a slice at $z$ = \(33\). AFNI~\cite{Cox1996} images showing brain activation from \textsc{Grad-CAM}.}
    \label{fig:dysvisual}
\end{figure*}

\begin{figure*}[t]
    \centering
    \includegraphics[width=.7\textwidth,height=4.5cm]{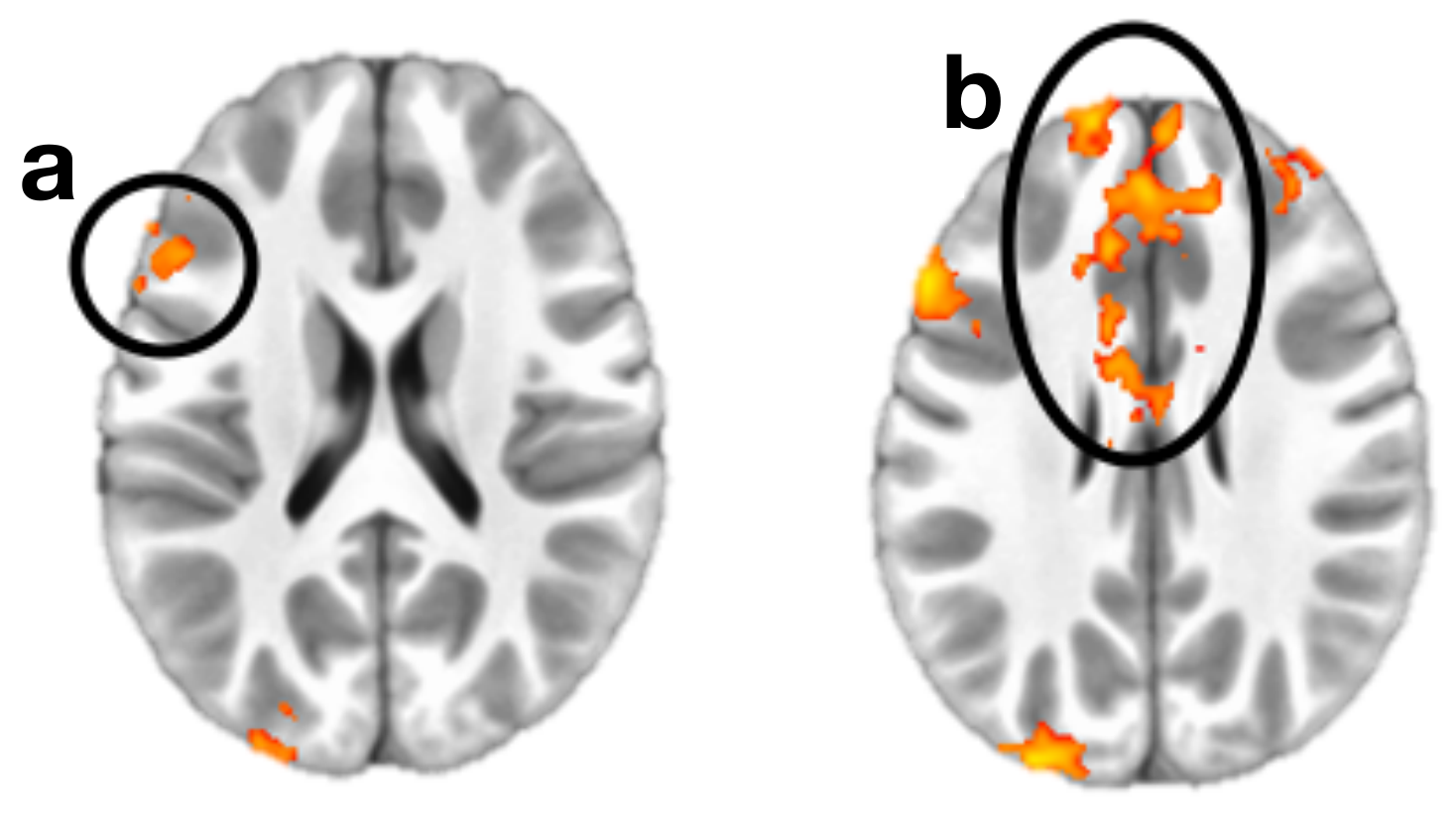}
    \caption{Example visual explanation for Typical readers subjects. Circle highlights instrumental brain regions for Dyslexic identification summarized in Table~\ref{tab:voxelcounttyp}. The left side of the images represent the left side of the brain. Image $a$ depicts a slice at $z$ = \(21\) and image $b$ depicts a slice at $z$ = \(27\). AFNI~\cite{Cox1996} images showing brain activation from \textsc{Grad-CAM}.}
    \label{fig:typvisual}
\end{figure*}





\begin{table*}
\begin{tabularx}{0.763\textwidth}{|l|l|l|l|l|l|}
\hline
  &  & \textbf{Regions} & \textbf{$x$} & \textbf{$y$} & \textbf{$z$} \\ \hline
Dyslexic & Left & Fusiform Gyrus & $-36$ & $-75$ & $-12$ \\ 
& & Precuneus & $-13$ & $-54$ & $16$ \\ 
& & Cuneus & $-10$ & $-63$ & $15$ \\
& & Isthmus & $-8$ & $-55$ & $16$ \\
& & Pars Opercularis & $-38$ & $2$ & $10$ \\
& & Transversal & $-41$ & $-16$ & $10$ \\ 
& & Precentral & $-53$ & $2$ & $10$ \\ \cline{2-6}
& Right & Rostral & $36$ & $44$ & $15$ \\ 
& & Pars Triangularis & $41$ & $37$ & $15$ \\ 
& & Superior Frontal & $3$ & $64$ & $15$\\ 
& & Insula & $37$ & $-3$ & $11$ \\ \cline{1-6}
Typical Readers & Left & Superior Frontal & $0$ & $41$ & $28$ \\ 
& & Caudate & $-8$ & $14$ & $27$ \\ 
& & Pars Opercularis & $-50$ & $20$ & $28$ \\ \cline{2-6} 
& Right & Pars Opercularis & $49$ & $19$ & $28$ \\ 
& & Rostral & $30$ & $39$ & $24$ \\ \cline{1-6}  
\end{tabularx}
\caption{Brain Regions Instrumental for Dyslexic and Typical Readers Identification with \textsc{Grad-CAM}~\cite{selvaraju2017grad}. Region labels follow Haskins pediatric atlas~\cite{molfese2015haskins}.}
\label{tab:brainregions}
\end{table*}

\section{Discussion and Related Work}

To our knowledge, there is little work on visual explanations and brain imaging; for instance, a recent study used these explanations for Alzheimer's disease(AD) and structural MRI (sMRI)~\cite{jin2019attention}. 
However, few approaches employed a visualization technique for MRI data, and there are none for fMRI data. 
The lack of approaches using brain imaging data of Dyslexia led us to search for related work employing deep learning to process any type of MRI data.    
Table~\ref{tab:related-work} summarizes previous work that employed deep learning~\cite{sarraf16,heinsfeld2018identification, jin2019attention} for subject classification, and approaches that applied machine learning to identify participants with dyslexia ~\cite{cui2016disrupted, tamboer2016machine, plonski2017multi}. 


The machine learning techniques we use in this article allow us to divide the related work into two types: i) work that aimed to identify participants with dyslexia using traditional machine learning algorithms (e.g. SVM); and ii) work that used Deep Neural Networks (DNNs) in brain imaging data for disease classification, as follows. 
\citet{sarraf16} employed the LeNet\hyp{5} architecture to classify patients with Alzheimer's disease.
\citet{heinsfeld2018identification} used two stacked denoising autoencoders for the unsupervised pre\hyp{training} stage to extract a lower\hyp{dimensional} version of the ABIDE (Autism Brain Imaging Data Exchange) data. 
\citet{jin2019attention} employed an attention-based 3D residual network based on the 3D ResNet to classify Alzheimer's Disease classification and to identify important regions in their visual explanation task. 
The remaining work applied machine learning techniques to classify dyslexic and control subjects. 
\citet{tamboer2016machine}, and \citet{cui2016disrupted} used SVM. 
\citet{plonski2017multi} on top of using SVM, also used logistic regression (LR), and random forest (RF).

Approaches that adopt deep learning models~\cite{sarraf16, heinsfeld2018identification, jin2019attention} show that DNN approaches can achieve competitive results using MRI and fMRI data.  
\citet{heinsfeld2018identification} achieved state-of-the-art results with 70\% accuracy in identification of ASD versus control patients in the dataset. 
The authors that used classic machine learning techniques~\cite{tamboer2016machine, cui2016disrupted, plonski2017multi} achieved $80\%$, $83.6\%$, and $65\%$ accuracy respectively on dyslexia prediction from anatomical scans. 
Performance of our deep learning models was consistent with other deep learning approaches for classification of neurological conditions. 
By contrast, our SVM results did not generalize as well as others~\cite{cui2016disrupted, tamboer2016machine, plonski2017multi}, but still outperformed another application of SVM for dyslexia classification~\cite{plonski2017multi}. 
Given the difference in datasets, we could not compare our approaches more directly. 

\citet{jin2019attention} visual explanations consisted of an attention map (much like a heatmap in visual representation) that indicated the significance of brain regions for AD classification. 
The authors compared their explanations to those generated by \textsc{3D\hyp{CAM}} and \textsc{3D\hyp{GRAD}\hyp{CAM}}~\cite{yang2018visual} methods. 
They~\cite{jin2019attention} observed that these two 3D methods led to a substantial drop in model performance when classifying subjects for Alzheimer's Disease by the extra calculations needed to generate the heatmaps.
By introducing the attention method, the authors obtained a 3D attention map for each testing sample and were able to identify the significance of brain regions related to changes in gray matter for AD classification. 
Our visualization technique may not be comparable to \citet{jin2019attention}, but the application of visualization techniques to medical imaging holds promise for making deep learning models interpretable.  

\begin{table*}
\centering
\begin{threeparttable}
\begin{tabularx}{\textwidth}{L m n m o m} 
\toprule
Study Reference & Modality & Dataset & Classifier & Task & Accuracy \\
\midrule
\textbf{Proposed Method} & Task based fMRI & ACERTA project & 2D CNN & Subject classification for Dyslexia &  \(94.83\%\)\\
\citet{sarraf16}  & rs-fMRI & ADNI\tnote{3} & LeNet-5 & Subject classification for Alzheimer & \(96.86\%\)\\ 
\citet{jin2019attention} & sMRI & ADNI\tnote{3} & Attention-based 3D ResNet & Subject classification for Alzheimer's Disease & \(92.1\%\) \\ 
\citet{cui2016disrupted} & sMRI & Private dataset & SVM & Subject classification for Dyslexia & \(83.6\%\)\\ 
\citet{tamboer2016machine} & sMRI & Non-disclosed dataset & SVM & Subject classification for Dyslexia & \(80\%\)\\ 
\citet{heinsfeld2018identification} & rs-fMRI & ABIDE & Denoising Autoencoder & Subject classification for Autism Spectrum Disorder & Above \(70\%\)\\ 
\citet{plonski2017multi} & sMRI & Private dataset & SVM, LR, RF & Subject classification for Dyslexia & \(65\%\) \\ 

\bottomrule
\end{tabularx}
\begin{tablenotes}\footnotesize
\item [1] Dataset:  https://predict-hd.lab.uiowa.edu/ \\
\item [2] fcon\_1000.projects.nitrc.org/indi/retro/cobre.htm \\
\item [3] adni.loni.usc.edu
\end{tablenotes}
\end{threeparttable}
\caption{Comparison with the classification scores of related work.} 
\label{tab:related-work}
\end{table*}

\section{Conclusion}

We introduce a novel approach for the investigation of neural patterns in task\hyp{based} fMRI that allow for the classification of dyslexic and control readers. 
While deep learning classifiers provide accurate identification of dyslexic versus control children based solely on their brain activation, such models are often hard to interpret.
In this context, our main contribution is a visualization technique of the features that lead to specific classifications, which allows neuroscience domain experts to interpret the resulting  models. 
Visual explanations of deep learning models allows us to compare regions instrumental to the classification with the latest neuroscientific evidence about dyslexia and the brain. 
The left occipital and inferior parietal regions that discriminated among groups are part of brain networks associated with phonological and lexical (word-level) processes in reading in different languages~\cite{paulesu2001dyslexia}. Other regions reported in the visualization are also associated with reading and reading disorders. More activation of anterior right-hemisphere prefrontal regions (e.g. right pars triangularis) are associated with dyslexia and possible compensatory mechanisms \cite{Vellutino2004, Shaywitz2005}.

Feature visualization techniques and visual explanations for deep learning models are a novel research area, and applying these techniques to neuroimaging data has the potential to help neuroscience research. 
Our work offers encouraging results, since the brain areas identified by the visual explanations are consistent with neuroscientific knowledge about the neural correlates of dyslexia. 
Nevertheless, there are a number of ways in which we can extend our work.
The deep learning classification models can be applied to publicly available, large fMRI or MRI datasets to investigate the areas that are instrumental for identification of, for example, autism spectrum disorder. 
Moreover, other visualization techniques can be applied to provide a qualitative comparison among  techniques when used to illustrate machine learning and deep learning studies of brain imaging.


%
%


%
%

\bibliographystyle{spbasic}      
\bibliography{refs}   

%
%

\end{document}